\pdfoutput=1 

\documentclass[final,authoryear,3p,times,twocolumn]{elsarticle}

\usepackage{amssymb}

\usepackage{bm}
\usepackage{color}

\newcommand{\e}[1]{\times 10^{#1}}

\usepackage{ulem}

\newcommand{\fig}[1]{Fig. \ref{#1}}
\newcommand{\eqn}[1]{Eqn. \ref{#1}}

\journal{New Astronomy}

\begin{document}

\begin{frontmatter}



\title{Accretion onto Stars with Octupole Magnetic Fields: Matter Flow, Hot Spots and Phase Shifts}


\author[1]{Min Long}
\author[2]{Marina M. Romanova}
\author[3,4]{Frederick K. Lamb}

\address[1]{Center for Theoretical Astrophysics, Department of Physics, University of Illinois at
Urbana-Champaigne, 1110 W Green St., Urbana, IL 61801, email: longm@illinois.edu}
\address[2]{Department of Astronomy, Cornell University, Ithaca, NY 14853, email:romanova@astro.cornell.edu}
\address[3]{Center for Theoretical Astrophysics, Department of Physics, University of Illinois at
Urbana-Champaigne, 1110 W Green St., Urbana, IL 61801, email: fkl@illinois.edu}
\address[4]{Also Department of Astronomy}

\begin{abstract}

Recent measurements of the surface magnetic fields of classical T~Tauri stars (CTTSs)
and magnetic cataclysmic variables show that their magnetic fields have a complex
structure. Investigation of accretion onto such stars requires global
three-dimensional (3D) magnetohydrodynamic (MHD) simulations, where the complexity of
simulations strongly increases with each higher-order multipole. Previously, we were
able to model disc accretion onto stars with magnetic fields described by a
superposition of dipole and quadrupole moments. However, in some stars, like  CTTS
V2129~Oph and BP Tau, the octupolar component is significant and it was necessary to
include the next octupolar component. Here, we show results of global 3D MHD
simulations of accretion onto stars with superposition of the dipole and octupole
fields, where we vary the ratio between components. Simulations show that if octupolar
field strongly dominates at the disc-magnetosphere boundary, then matter flows into
the ring-like octupolar poles, forming ring-shape spots at the surface of the star
above and below equator. The light-curves are complex and may have two peaks per
period. In case where the dipole field dominates, matter accretes in two ordered
funnel streams towards poles, however the polar spots are meridionally-elongated due
to the action of the octupolar component. In the case when the fields are of similar
strengths,
 both, polar and belt-like spots are present. In many cases the light-curves show
 the evidence of complex fields, excluding the cases of small inclinations angles, where
sinusoidal light-curve is observed and `hides' the information
about the field complexity.

We also propose new mechanisms of phase shift in stars
with complex magnetic fields.  We suggest that the phase shifts
can be connected with: (1) temporal variation of the star's
intrinsic magnetic field and subsequent redistribution of main
magnetic poles;  (2) variation of the accretion rate, which causes
the disc to interact with the magnetic fields associated with
different magnetic moments. We use our model to
demonstrate these phase shift mechanisms, and we discuss possible
applications of these mechanisms to accreting millisecond pulsars
and young stars.

\end{abstract}


\end{frontmatter}


\label{}

\section{Introduction}

Different types of stars may have dynamically important
magnetic fields. These are young classical T Tauri
stars (CTTSs) (Basri, Marcy \& Valenti 1992; Johns-Krull 2007),
magnetic white dwarfs in some cataclysmic variables
(e.g., Warner 1995; Euchner et al. 2005), and various types of
neutron stars
(e.g., \citealt{maki99}, van der Klis 2000; Ghosh 2007).
In most stars the structure of the field is unknown.

Measurements of the surface magnetic fields of CTTSs using
different techniques indicate that they have a complex structure
(Johns-Krull, Valenti \& Koresko 1999; Johns-Krull 2007).
Measurements of the magnetic fields of nearby low-mass stars using
the Zeeman-Doppler technique show that their fields are often
complex (e.g., \citealt{dona97},  Donati et al.\ 1999;
\citealt{jard02}).
The observed surface field is often
approximated with a set of multipoles whose magnetic moments have
different misalignment angles relative to the rotational axis
(i.e., different tilts) and different phases in the longitudinal
direction (e.g., \citealt{jard02}). Recent observations of two
CTTSs have shown that in one star (V2129 Oph) the surface magnetic
field can be approximated with the superposition of a
slightly (but differently) tilted octupole, which dominates, and a
dipole, where the magnetic moments are oriented at different
phases (Donati et al.\ 2007, 2011). In the other star, BP
Tau, the magnetic field can be approximated with slightly (but
differently) tilted dipole and octupole moments of equal
strengths, where the magnetic moments are in anti-phase (Donati et
al.\ 2008). Other multipoles contain much less magnetic energy. An
example of these young stars shows that such a distribution might
be common for some other magnetized stars.

Recently, the magnetic field structure of young stars with complex
fields has been studied analytically. It was found that the
fraction of the flux through the stellar surface in open field
lines is smaller in stars with complex fields than in those with
purely dipolar fields (Gregory et al.\ 2008) and a superposition
of multipole fields yields smaller accretion hot spots (Mohanty \&
Shu 2008). The potential approximation is usually used to
extrapolate the surface magnetic field to larger distances.
That is, it is suggested that the field outside of the
star is not disturbed by the surrounding plasma (e.g.,
\citealt{greg10}). Gregory et al.\ (2006) calculated
possible paths of gas flow around stars with magnetic
fields constructed from measurements using the potential
approximation. Recently, we were able to compute the flow of gas
around V2129~Oph and BP~Tau in global 3D MHD simulations in which
the magnetic field is not assumed to be a potential field but is
instead calculated as part of the simulation (Romanova et al.\
2011a; Long et al. \ 2011).

Zeeman tomography of magnetic white dwarfs has shown that they
also may have complex magnetic fields (Euchner et al.\ 2002). In
some, such as HE~1045$-$0908, the magnetic field associated with
the star's quadrupole moment dominates whereas the fields
associated with its dipole and octupole moments are much weaker
(Euchner et al.\ 2005). In others, representations of the field
require inclusion of tilted dipole, quadrupole, octupole, and
other multipoles up to $n=4$ or $n=5$ (Euchner et al.\ 2006;
Beuermann et al.\ 2007). These results are in accord with earlier
indications of field complexity, such as asymmetric distribution
of spots on the stellar surface (Meggitt \& Wickramasinghe 1989;
Piirola et al.\ 1987; see Wickramasinghe \& Ferrario 2000 for a
review).

Many neutron stars also have dynamically important
magnetic fields. X-ray observations of the cyclotron line
helped to reveal the value of the field in some neutron stars
(e.g., \citealt{maki99,cobu02} and references therein; see also
Bignami et al. 2003).  The magnetic fields of newly formed neutron
stars may be enhanced by a dynamo mechanism (e.g., Thompson \&
Duncan 1993) and may have a complex structure. There is evidence
that at least some accretion-powered X-ray pulsars have complex
magnetic field structures (Elsner \& Lamb 1976;  Burderi et al,
2000;  Nishimura 2005). Accreting millisecond X-ray
pulsars have relatively weak magnetic fields (e.g. Psaltis \&
Chakrabarty 1999; van der Klis 2000).  They often have almost
sinusoidal light curves, or the light-curve can strongly depart
from a sinusoid during the outburst and may have two peaks per
period (see, e.g., Hartman et al. 2008; Poutanen, Ibragimov \&
Annala 2009; Ibragimov \& Poutanen 2009). A sinusoidal light-curve
is consistent with a dipole magnetic field with a small tilt to
the spin axis. However, in stars with more complex fields, the
light-curve can be also sinusoidal or almost sinusoidal, in
particular at small inclination angles (e.g., Long et al. 2007,
2008; see this paper). Also, the field may be a dipole at large
distances, but have more complex structure at small distances. In
some theories, it is suggested that the field of millisecond
pulsars could represent a strongly off-center dipole field (e.g.,
Ruderman 1991; Chen \& Ruderman 1993), which gives us a hint that
the field may be more complex than a typical dipole field.

An unusual phenomenon
--- rapid, relatively large shifts in the phase of the light curve
--- have been observed in many millisecond pulsars (e.g., Morgan
et al.\ 2003; Markwardt 2004; Burderi et al.\ 2006; Hartman et
al.\ 2008, 2009). There may be different reasons for the phase
shifts. In one model, the phase shifts are connected with the
motion of accretion spots on the surface of the star (e.g., Lamb
et al. 2009a,b; Patruno et al. 2010). Such spot motion has been
observed in global 3D simulations of disc-magnetosphere
interactions (Romanova et al. 2003, 2004; Bachetti et al. 2010).
In another model, it is suggested that the timing noise can be
connected with variation of the pulse shapes
 (e.g., Poutanen et al. 2009). This problem has not been solved yet.

In this paper, we propose and investigate different types
of phase shifts that can be connected with the \textit{complexity}
of stellar magnetic fields. In one mechanism, we take into account
that the magnetic field of the star may vary in time, and hence
the position of hot spots will vary as well; this will be observed
as phase-shift of pulses. The magnetic field of young, low-mass
stars may vary rapidly due to the dynamo mechanism.
Spectropolarimetric observations of young CTTS stars, show that in
some stars, the magnetic field can vary as fast as one week
(Donati et al. 2010; see also \citealt{smir04}). This mechanism is
evidently important in young stars, where the magnetic field can
vary rapidly; the magnetic field of compact stars varies only
slowly (e.g., Ruderman 2005). It is possible, however, that the
field can be varied or partially buried during periods of enhanced
accretion, as we discuss later in this paper (see Sec.
\ref{sec:timing-noise}).

Another type of phase-shift can be connected with
variation of the accretion rate, and it operates even in cases
when the magnetic field is fixed. Namely, if the accretion rate
varies, then the accretion disk interacts with different multipoles:
at low accretion rates it will interact with the dipole
component, which dominates at large distances; at higher accretion
rates, the disc will come closer to the star and it will interact
with the higher-order multipole of the complex field. The magnetic
axes of multipoles can have different tilts and phases relative to
one another, and hence the set of hot spots and the light-curves
will be different at different accretion rates. We can expect
variation in both the phases and the shape of the light-curves. We use
our present stellar model with a superposition of the dipole and
octupole fields to demonstrate both of the above models of phase-shifts.

3D simulations of accretion onto stars with multipolar fields are
very challenging due to the steep gradients of the magnetic field.
The complexity strongly increases with adding higher order
multipoles. For example, in the case of the dipole field, the
magnetic energy terms are on the order of $B_1^2\sim 1/r^6$. The
code should resolve the gradients of this value in the vicinity of
the star. In the case of a quadrupole field, we deal with terms on
the order of $B_1^2\sim 1/r^8$, and in the case of octupolar field
with terms $\sim 1/r^{10}$.

In our earlier studies, we were able to perform global 3D~MHD
simulations of accretion onto stars with a dipole magnetic field
(Koldoba et al.\ 2002; Romanova et al.\ 2003, 2004; Kulkarni \&
Romanova 2005) and  stars with a combination of dipole and
quadrupole fields in the cases when they are aligned (Long, et al.
2007) or misaligned (Long, et al. 2008).

In this paper, we investigate accretion onto stars with an {\it
octupole} field component. Our methods allow us to investigate the
general case of a magnetic field produced by superposition of
dipole, quadrupole, and octupole fields oriented in different
directions. However, even the simpler case of a magnetic field
produced by superposition of tilted dipole and quadrupole fields
usually creates  a very complex field structure which is difficult
to analyze (e.g., Long et al.\ 2007, 2008). Consequently, for
clarity we omit quadrupole fields in this work, restricting our
consideration to the clearer cases of a superposition of octupole
and dipole fields,in order to focus on the role the octupole field
plays in influencing the matter flow, the shapes of the resulting
hot spots, and the light curves produced by these hot spots.

In separate papers, we investigated accretion onto CTTSs V2129 Oph
and BP Tau, where the  magnetic field configuration has been
obtained from observations and is fixed (Romanova et al. 2011a;
Long et al. 2011). In both cases, the octupolar component is not
very strong relative to the dipole component. In this paper, we
perform a systematic study of different combinations of the dipole
and octupole components. In particular, we show our result for the
case when the octupolar component  strongly dominates.

Section 2 describes the simulation method and the magnetic field
configurations. Section~3 presents our results for different
combinations of dipole and octupole fields. Section~4 is devoted
to the analysis of phase shifts in stars with complex fields.
Section~5 summarizes and discusses our results.

\section{Numerical model and magnetic field configurations}

\subsection{Model}

\begin{figure}
\centering
\includegraphics[scale=1.0]{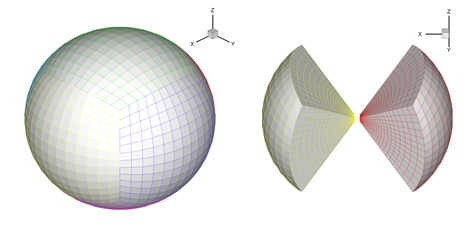}
\caption{\label{grid} A low resolution ``cubed sphere" grid
 is shown for demonstration. The left
panel shows the grid on the surface of the inflated cube. Each of
the six cube sides has $N^2=13^2$ curvilinear Cartesian grid
cells. The grid consists of  $N_r=28$ concentric spheres. The
right panel shows  two sectors of the grid. Our simulations use a
higher grid resolution, up to $N_r\times N^2=200\times 51^2$ in
each sector.}
\end{figure}

\begin{figure}
\centering
\includegraphics[scale=0.3]{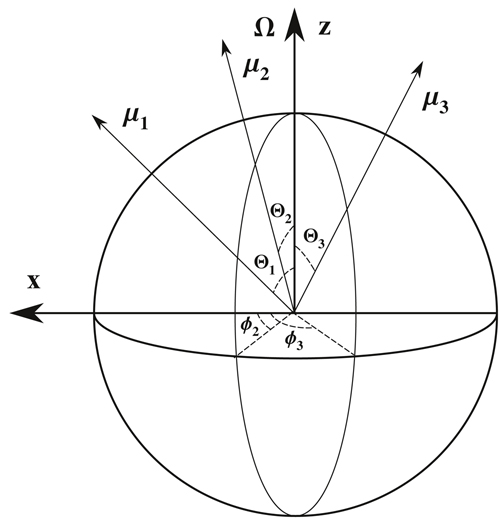}
\caption{\label{config} Sketch of the directions of the multipole
magnetic moments. The rotation axis $\bm{\Omega}$ is in the $z$
direction.  The magnetic axes of the dipole $\bm{\mu_1}$,
quadrupole $\bm{\mu_2}$ and octupole $\bm{\mu_3}$ are inclined at
angles of $\Theta_1$, $\Theta_2$ and $\Theta_3$ from the $z-$
axis. The dipole moment $\bm\mu_1$ is placed in the $xz$ plane.
The angles between $\bm{\Omega-\mu_2}$, $\bm{\Omega-\mu_3}$ and
$xz$ planes are $\phi_2$ and $\phi_3$ , respectively.}
\end{figure}

Our 3D MHD model has been described in a series of papers (Koldoba
et al.\ 2002; Romanova et al.\ 2003; 2004; Kulkarni \& Romanova
2005; Long et al. 2007, 2008), where the disc accretion onto stars
with dipole and more complex magnetic fields has been
investigated. Here, we briefly summarize the different aspects of
the model.

\subsubsection{Dimensionless Variables and Reference Units}\label{ref-units}

The MHD equations are solved using dimensionless variables
$\widetilde A$. To obtain the physical dimensional values $A$, the
dimensionless values $\widetilde{A}$ should be multiplied by the
corresponding reference units $A_0$; $A=\widetilde{A}A_0$. To
choose the reference units, we first choose the stellar mass
$M_\star$ and radius $R_\star$. The reference units are then
chosen as follows: mass $M_0=M_\star$, distance
$R_0=R_\star/0.35$, velocity $v_0=(GM_0/R_0)^{1/2}$, time scale
$P_0=2\pi R_0/v_0$, angular velocity $\Omega_0=v_0/R_0$. The
reference magnetic field $B_0$ can be obtained by choosing a
reasonable fiducial value for the surface dipole field strength
$B_{1\star}$. Then $B_0$ is the fiducial dipole field strength at
$R_0$; $B_0=B_{1\star }(R_\star/R_0)^3$. We then define the
reference dipole moment $\mu_{1,0}=B_0R_0^3$, quadrupole moment
$\mu_{2,0}=B_0R_0^4$, octupole moment ${\mu_{3,0}}=B_0R_0^5$,
density $\rho_0=B_0^2/v_0^2$, pressure $p_0=\rho_0v_0^2$, mass
accretion rate $\dot{M}_0=\rho_0v_0R_0^2$, angular momentum flux
$\dot{L}_0=\rho_0v_0^2R_0^3$, energy per unit time
$\dot{E}_0=\rho_0v_0^3R_0^2$ (the radiation flux $J$ is also in
units of $\dot{E}_0$), temperature $T_0=\mathcal{R}p_0/\rho_0$,
where $\mathcal{R}$ is the gas constant, and the effective
blackbody temperature $T_{\mathrm{eff,0}} = (\rho_0
v_0^3/\sigma)^{1/4}$, where $\sigma$ is the Stefan-Boltzmann
constant.

Therefore, the dimensionless variables are $\widetilde{r}=r/R_0$,
$\widetilde{v}=v/v_0$, $\widetilde{t}=t/P_0$,
$\widetilde{B}_n=B_n/B_0$, $\widetilde{\mu}_n = \mu_n / \mu_{n,0}$
($n=1,2,3$ for dipole, quadrupole and octupole components) and so
on. In the subsequent sections, we show dimensionless values for
all quantities and drop the tildes ($\sim$). Our dimensionless
simulations are applicable to different astrophysical objects with
different scales. We list the reference values for typical CTTSs,
cataclysmic variables, and millisecond pulsars in Tab.
\ref{tab:refval}.

\begin{table}
\centering
\begin{tabular}{l@{\extracolsep{0.2em}}l@{}lll}

\hline
& CTTSs       & White dwarfs          & Neutron stars           \\
\hline

{$M_\star(M_\odot)$}              & 0.8         & 1                     & 1.4                     \\
{$R_\star$}                       & $2R_\odot$  & 5000 km               & 10 km                   \\
{$B_{1\star}$ (G)}                & $10^3$      & $10^6$                & $10^9$                  \\
{$R_0$ (cm)}                      & $4\e{11}$   & $1.4\e9$              & $2.9\e6$                \\
{$v_0$ (cm s$^{-1}$)}             & $1.6\e7$    & $3\e8$                & $8.1\e9$                \\
{$\Omega_0$ (s$^{-1}$)}           & $4\e{-5}$   & 0.2                   & $2.8\e3$                \\
{$P_0$}                           & $1.8$ days  & 29 s                  & 2.2 ms                  \\
{$B_0$ (G)}                       & 43          & $4.3\e4$              & $4.3\e7$                \\
{$\rho_0$ (g cm$^{-3}$)}          & $7\e{-12}$  & $2\e{-8}$             & $2.8\e{-5}$             \\
{$p_0$ (dy cm$^{-2}$)}            & $1.8\e{3}$  & $1.8\e{9}$            & $1.8\e{15}$             \\
{$\dot M_0(M_\odot$yr$^{-1})$}    & $2.8\e{-7}$ & $1.9\e{-7}$           & $2.9\e{-8}$             \\
{$\dot{L}_0$ (g cm$^2$s$^{-2}$)}  & $1.15\e{38}$& $4.9\e{36}$           & $4.5\e{34}$             \\
{$T_0$ (K)}                       & $1.6\e6$    & $5.6\e8$              & $3.9\e{11}$             \\
{$\dot E_0$ (erg s$^{-1}$)}       & $4.8\e{33}$ & $1.2\e{36}$           & $1.2\e{38}$             \\
{$T_{\mathrm{eff},0}$ (K)}        & 4800        & $3.2\e5$              & $2.3\e7$                \\
\hline
\end{tabular}
\caption{Sample reference units for typical CTTSs, cataclysmic
variables, and millisecond pulsars. Real dimensional values for
variables can be obtained by multiplying the dimensionless values
of variables by these reference units.} \label{tab:refval}
\end{table}

\subsubsection{Initial Conditions}
We consider a rotating magnetic star surrounded by an accretion
disc and a corona. The disc is cold and dense, while the corona is
hot and rarefied, and at the reference point ( the inner edge of
the disc in the disc plane ), $T_c=100T_d$, $\rho_c=0.01\rho_d$,
where the subscripts `d' and `c' denote the disc and the corona.
The disc and corona are initially in rotational hydrodynamic
equilibrium, where the sum of the gravitational, centrifugal, and
pressure gradient forces is zero at each point in the simulation
region. To avoid initial magnetic field discontinuity at the
disc-corona boundary, the corona is set to rotate with the
Keplerian angular velocity at each cylindrical radius $r$. An
$\alpha$-type viscosity is incorporated into the code. It operates
only in the disc (above some density level, $\rho\gtrsim 0.2$) and
helps regulate the accretion rate. We use a small value of
$\alpha=0.02$ for the majority of our simulation runs, and we use
$\alpha=0.04$ for the cases where the octupole
field strongly dominates. \footnote{Note that the cases
with dominantly octupolar fields are the most difficult to model, and
the simulations do not last long. Somewhat faster-accreting disks
helped to simulate these difficult cases over longer time scales.}
Other parameters of the disc such as the density distribution and
initial structure are fixed in all of the simulations.

\begin{figure}
\centering
\includegraphics[scale=1.2]{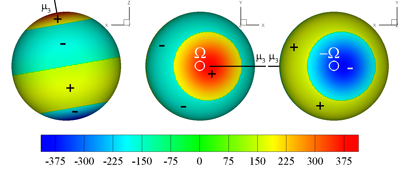}
\caption{\label{bsurfoct} The magnetic field distribution on the
star with an octupole field, $\mu_3=0.2$, $\Theta_{3}=5^\circ$, as
seen from the equatorial plane (left panel), the north pole
(middle panel) and the south pole (right panel). }
\end{figure}

\begin{figure}
\centering
\includegraphics[scale=1.1]{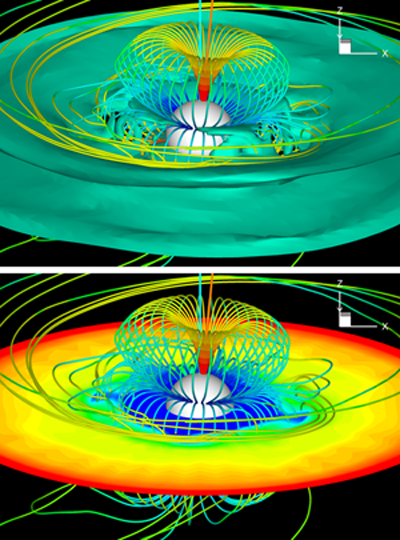}
\caption{\label{3doct} 3D views of matter flow to a star with an
octupolar field with $\mu_3=0.2$, $\Theta_3=5^\circ$ at $t=4.5$.
The disc is shown by a constant density surface in green with
$\rho=0.15$ in the top panel. Different density levels in the
equatorial plane are shown in the bottom panel. The colors along
the field lines represent different polarities and strengths of
the field. The thick cyan and orange lines represent the rotation
and octupole moment axes, respectively. }
\end{figure}

\subsubsection{Boundary Conditions}
At the inner boundary (the surface of the star), most $A$
variables are set to have free boundary conditions, ${\partial
A}/{\partial r}=0$. The initial magnetic field on the surface of
the star is taken to be a superposition of tilted dipole and
octupole fields (see $B_1$ and $B_3$ in \eqn{e1}). As the
simulation proceeds, we assume that the normal components of the
fields remain unchanged, i.e., the magnetic field is frozen at the
surface of the star. We neglect possible changes in the magnetic
field structure inside the star due to, e.g., dynamo processes,
because in most cases the time scale of such variations is much
longer than the length of the simulation time. Our
simulations last for $5-50$ Keplerian rotations at the inner disc;
the longest runs correspond to the dipole-dominated cases, while
the shortest run corresponds to a purely octupolar case, which is
numerically challenging.

The outer boundary is located at a radius of
$r_{max}\approx 36 R_\star$. The simulation region is large enough
 and the disc is massive enough to
supply matter for the entire duration of the simulations. At the
outer boundary, free conditions are taken for all variables. In
addition, matter is not permitted to flow back from the
outer boundary back into the simulation region.

\subsubsection{The ``Cubed sphere" Grid}

The 3D MHD equations are solved with a Godunov-type code on the
``cubed sphere" grid (Koldoba et al.\ 2002; see also Putman \& Lin
2007). The grid consists of $N_r$ concentric spheres, where each
sphere represents an inflated cube. \fig{grid} shows that the grid
consists of six sectors corresponding to six sides of the cube
with $N\times N$ curvilinear Cartesian grids on each side. The
whole grid consists of $6\times N_r\times N^2$ cells. For modeling
the octupole fields, a higher radial grid resolution is needed
near the star compared with the pure dipole cases. We accomplish
this by choosing the radial size of the grid cells to be 2.5 times
smaller than the angular size in the region $r < (4-6)R_\star$
(see \S\ref{ref-units}), while it is equal to the angular size in
the outer region as in all our previous work. The typical grid
used in simulations has $6\times N_r\times N^2 =
6\times200\times51^2$ grid cells. Simulations with higher/lower
grid resolutions were performed for comparisons.

\begin{figure}
\centering
\includegraphics[scale=1.0]{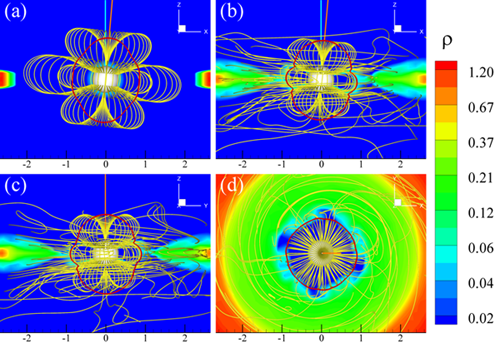}
\caption{\label{flowoct} The density distribution in different
planes (color background) and 3D magnetic field lines (yellow
lines) for an octupole configuration with $\mu_3=0.2$,
$\Theta_{3}=5^\circ$. Panel (a) shows an $xz$ slice at $t=0$;
panels (b), (c) and (d) show $xz$, $yz$ and $xy$ slices at
$t=4.5$. The red lines show surfaces where $\beta_1=1$. The thick
cyan and orange lines represent the rotation and octupole moment
axes respectively.}
\end{figure}

\subsection{Magnetic Field Configurations}

When electrical currents outside the star can be neglected, the magnetic
field there can be described by a magnetic
scalar potential $\varphi (\mathbf{r})$ and can be represented as a sum of multipoles.
The first three multipoles can be written as
\begin{equation}
\label{e0} \mathbf{B(r)}=\mathbf{B_1}+\mathbf{B_2}+\mathbf{B_3}\;,
\end{equation}
where
\begin{eqnarray}
\label{e1} \mathbf{B_{1}}&=&\frac{3\mu_1(\hat{\bm\mu_1}\cdot{\hat{\bf r}})}
{r^3}\hat{\bf r}-\frac{\mu_1}{r^3}\hat{\bm\mu_1}\nonumber\\
\mathbf{B_{2}}&=&\frac{3\mu_2}{4r^4}(5(\hat{\bm\mu_2}\cdot\hat{\bf r}) ^2-1)\hat{\bf r}-\frac{3\mu_2}{2r^4}(\hat{\bm\mu_2}\cdot\hat{\bf r})
\hat{\bm\mu_2}\nonumber\\
\mathbf{B_{3}}&=&5\big(\frac{\mu_3}{2r^5}\big)(\hat{\bm\mu_3}\cdot\hat {\bf r})\big[7(\hat{\bm\mu_3}\cdot\hat{\bf r})^2-3\big]\hat{\bf r}
\nonumber\\
&&-3\big(\frac{\mu_3}{2r^5}\big)\big[5(\hat{\bm\mu_3}\cdot\hat{\bf r}) ^2-1\big]\hat{\bm\mu_3}\;.
\end{eqnarray}
Here $\bf B_1$, $\bf B_2$, and $\bf B_3$ are the magnetic fields
produced by the symmetric dipole, quadrupole and octupole moments
and $ \hat{\bf r}$, $\hat{\bm\mu_1}$, $\hat{\bm\mu_2}$, and
$\hat{\bm\mu_3}$ are the unit vectors describing the position and
the symmetric dipole, quadrupole, and octupole moments,
respectively. The magnetic moments may be inclined at angles
$\Theta_{1}$, $\Theta_2$, and $\Theta_{3}$ relative to the
rotation axis $\bm{\Omega} $. They may also be in different
meridional planes $\bm{\Omega-\mu_2}$ and $\bm{\Omega-\mu_3}$,
with an azimuthal angles $\phi_2$ and $\phi_{3}$ relative to the
$\bm{\Omega- \mu_1}$ plane defined by the dipole moment and the
rotation axis. \fig{config} illustrates the geometry by an example
in which all three magnetic moments have different orientations.
For our simulations, we take only the dipolar and octupolar
components and investigate the role of the octupolar component.

\begin{figure}
\centering
\includegraphics[scale=1.2]{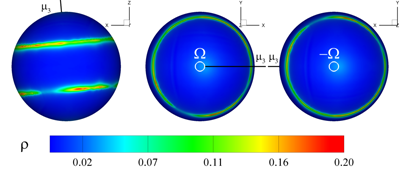}
\caption{\label{hsoct} The surface density distribution on the
star with an octupole magnetic field, $\mu_3=0.2$,
$\Theta_3=5^\circ$, as seen from the equatorial plane (left
panel), the north pole (middle panel), and the south pole (right
panel).}
\end{figure}


\subsection{The Alfv\'en Surface and Magnetospheric Radius}
The magnetospheric radius, $r_m$, is a characteristic radius where the inflowing matter is stopped by the
magnetosphere. Several expressions for this radius have been derived since the 1970s for stars with dipole fields. The
first estimations were done for spherically-accreting matter. It was suggested that the radially-falling matter is
stopped by the magnetic pressure of the magnetosphere (e.g., Lamb et al.\ 1973):
\begin{equation}\label{e11}
B(r_A)^2/8\pi=\rho(r_A)v(r_A)^2 ,
\end{equation}
\begin{equation}\label{e2}
r_A = kr_A^{(0)}, \ \ r_A^{(0)}=\dot{M}^{-2/7}\mu_1^{4/7}(GM)^{-1/7},
\end{equation}
where $r_A$ is the Alfv\'{e}n radius and $k$ is a dimensionless
coefficient on the order of unity (Elsner \& Lamb 1977; Ghosh et
al.\ 1977). For {\it disc} accretion, a similar criterion can be
used but for stresses. The disc rotates in the azimuthal direction
and hence main stresses are connected with the azimuthal
components of the stress tensor: $T_{\phi\phi}=[p + \rho v_\phi^2]
+ [B^2/8\pi-B_\phi^2/4\pi]$ (here we neglected the viscous stress
which is much smaller than the matter stress). The motion of the
disc is disturbed by the rotating magnetosphere, when the matter
stresses become comparable with the magnetic stresses, or $p +
\rho v_\phi^2 = B^2/8\pi-B_\phi^2/4\pi$. The dominant component of
the dipole magnetic field is the poloidal component, and hence
$B_\phi<<B$ and we obtain the condition for stresses as  $p + \rho
v_\phi^2 = B^2/8\pi$.

Different criteria were proposed by other authors (Collier Cameron \& Campbell 1993; Armitage \& Clarke 1996;  Matt \&
Pudritz 2005). Analysis of these criteria shows that most of them give a radius similar to that given by \eqn{e2}, but
with different coefficients $k$ (Bessolaz et al.\ 2008). A comparison of that formula with simulations gives
$k\approx0.5$ (Long, Romanova \& Lovelace 2005), which is very close to the coefficient estimated by Ghosh \& Lamb
(1979).

For multipolar fields, we use the generalized formula, such that the $n-$th component of the field is
$B_n\sim\mu_n/r^{n+2}$, (see details in Appendix \ref{rm-mult}), and the magnetospheric radius
\begin{equation}\label{e22}
r_m=k_n r_{m,n}^{(0)},~~~ r_{m,n}^{(0)}=\mu_n^{\frac{4}{4n+3}}\dot{M}^{-\frac{2}{4n+3}}(GM)^{-\frac{1}{4n+3}}.
\end{equation}

\begin{figure}
\centering
\includegraphics[scale=1.2]{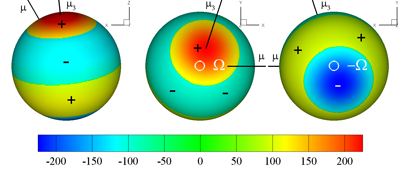}
\caption{\label{bsurfwdip} The magnetic field distribution on the
star with a strong octupole and a weak dipole component,
$\mu_1=0.2$, $\mu_3=0.3$, $\Theta_1=30^\circ$,
$\Theta_3=20^\circ$, $\phi_3=70^\circ$, as seen from the
equatorial plane (left panel), the north pole (middle panel) and
the south pole (right panel). }
\end{figure}

\begin{figure}
\centering
\includegraphics[scale=1.1]{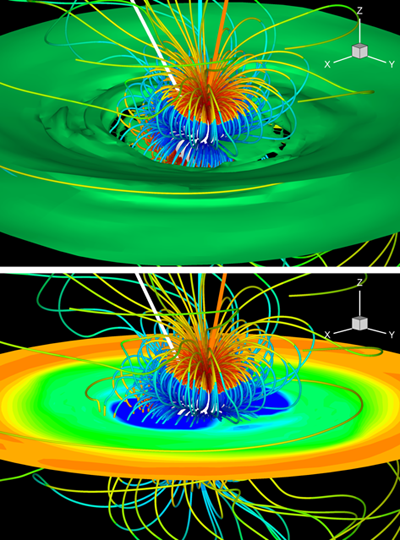}
\caption{\label{3dwdip} 3D views of matter flow to a star with a
strong octupole and weak dipole magnetic field, $\mu_1=0.2$,
$\mu_3=0.3$, $\Theta_1=30^\circ$, $\Theta_3=20^\circ$,
$\phi_3=70^\circ$, at $t=5$. The disc is shown by a constant
density surface in green with $\rho=0.2$ in the top panel;
different density levels in the equatorial plane are shown in the
bottom panel. The colors along the field lines represent different
polarities and strengths of the field. The thick cyan, white and
orange lines represent the rotation, dipole and octupole axes
respectively.}
\end{figure}

\section{Accretion onto Stars with an Octupole and a Dipole Field}

In all of our simulations, we take a disc  with the same
initial density and pressure distribution.  Then, we choose
magnetic configurations with different strengths of the dipole and
octupole fields, such that the disc is stopped either by the
octupolar component of the field (in octupole-dominated cases), by
the dipole component (in dipole-dominated cases), or by the dipole
and octupole components of comparable strengths.  We
discuss these cases below in detail.

\subsection{Pure octupole}

First, we consider accretion onto a star with an octupole magnetic field with $\mu_3=0.2$. The magnetic axis of the
octupole field is tilted relative to the stellar spin axis at a small angle, $\Theta_{3}=5^\circ$.

\fig{bsurfoct} shows the magnetic field distribution on the stellar surface. There are two polar regions with strong
positive (red) and negative (blue) polarities, as seen in the middle and right panels. There are also two octupolar
belts with opposite polarities as seen in the left panel. A meridian line which goes from the north to the south
magnetic poles will pass through the positive northern pole, negative northern belt, positive southern belt and
negative southern pole with zero magnetic field at the magnetic equator. This distribution differs from the dipole
field which has only positive and negative poles.

\fig{3doct} (top panel) shows 3D views of matter flow to the star at time $t=4.5$. The magnetic field near the star has
an octupolar shape with three sets of closed loops of field lines connecting regions of different polarities on the
star, which we call {\it northern, southern} and {\it equatorial} sets of loops. Some field lines are closed, while
others are dragged by the disc, wrapped around the rotation axis and inflate into the corona. Matter of the disc is
lifted above the equatorial set of loops and flows into the northern magnetic belt on the star.  The bottom panel shows
the density distribution in the equatorial plane, where the disc matter is stopped by the equatorial set of loops and
forms a low-density gap around the star.

\fig{flowoct} shows the density distribution in different planes
and sample 3D magnetic field lines.  Panel (a) shows that at $t=0$
the magnetic field is a pure octupole field, where the three sets
of loops of field lines are clearly seen. The equatorial set of
loops is expanded to larger distances than the polar sets, to
demonstrate the interaction of the magnetosphere with the disc.
Panels (b) and (c) show that the disc matter is stopped by the
magnetosphere, lifted above the disc plane and channeled by the
equatorial set of loops, before flowing into the octupolar belts
on the stellar surface. Panel (b) shows that matter flow is not
axisymmetric: more matter flows in from the direction towards
which the octupole is inclined. Panel (d) shows that there is a
low-density magnetospheric gap around the star.

The bold red line represents the magnetospheric surface,
 where the matter stresses equal
the magnetic stresses,
\begin{equation}
\label{e-beta} \beta_1=\frac{p+\rho v^2}{B^2/8\pi}=1.
\end{equation}
The magnetic field lines stay closed inside the magnetospheric surface, where the magnetic
stresses dominate.

The initial octupolar field shown in  panel (a) is a {\it
potential} field, that is, a field that has not been disturbed by
the surrounding plasma. The potential approximation is often used
in extrapolation of the observed surface fields to larger
distances (e.g., Jardine et al.\ 2002; Gregory et al.\ 2006;
Donati et al.\ 2007). Panels (b)-(d) show that the potential
approximation is sufficiently good inside the Alfv\'en surface,
where $\beta_1 < 1$. However, the field strongly departs from the
potential one at larger distances.

\fig{hsoct} shows that the hot spots on the stellar surface represent
 two rings which are located in the planes
parallel to the magnetic equator.  Each ring has a density enhancement
on one side, where disc accretion is higher due
to the inclination of the magnetic axis.  The middle and right panels
show that the density enhancements in the
northern and southern rings are antisymmetric relative to the
magnetic equatorial plane.

\subsection{Strong octupole and weak dipole}

Next, we consider the accretion onto a star with a superposition
of a relatively strong octupole and a much weaker dipole component
on the stellar surface: $\mu_1=0.2$, $\mu_3=0.3$,
$\Theta_1=30^\circ$, $\Theta_3=20^\circ$, $\phi_3=70^\circ$. Here
the rotation axis, dipole and octupole moments are all misaligned.

First, we estimate the radius at which the magnetic fields of the dipole and octupole components are equal. Using
\eqn{e1}, and suggesting for simplicity that the dipole and octupole moments are aligned, we obtain approximate
formulae for the equatorial and polar magnetic fields: (1) Equatorial field: $B_1=\mu_1/r^3$,  $B_3=(3/2)\mu_3/r^5$,
and (2) Polar field: $B_1=2\mu_1/r^3$, $B_3=4\mu_3/r^5$. By equating $B_1$ and $B_3$ we obtain the distances where the
dipole and octupole components are equal in the equatorial and polar directions:
\begin{equation}
\label{e-rcrit}
r_{eq}=\bigg({\frac{3}{2}\frac{\mu_3}{\mu_1}}\bigg)^\frac{1}{2}, ~~r_{pole}=\bigg({2 \frac{\mu_3}{\mu_1}}\bigg)^\frac{1}{2}.
\end{equation}
Substituting for $\mu_1$ and $\mu_3$, we obtain $r_{eq}=1.5$ and
$r_{pole}=1.4$. At smaller/larger distances the octupole/dipole
field dominates. It is evident that near the star, $r\sim
R_\star=0.35$, the octupolar component is much stronger than the
dipole component. That is why in \fig{bsurfwdip}, the magnetic
field distribution shows similar features to those in the pure
octupole case (\fig{bsurfoct}).

\fig{3dwdip} shows 3D views of accretion flow onto the star at
time $t=5$. The disc matter comes close to the star and mainly
interacts with the octupolar component and accretes onto the
octupolar belts on the stellar surface. External field lines are
dragged by the disc and inflate. This case shows similar features
to that of the pure octupole, because in both cases the octupole
determines matter flow. We should note that if the
accretion rate were much smaller, then the inner disk would
move to larger distances and the dipole component could interact
with the disk.  This case differs from that of a pure octupole
field with no dipole component.

\begin{figure}
\centering
\includegraphics[scale=1.0]{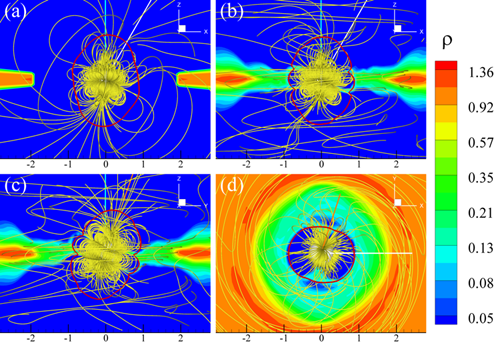}
\caption{\label{flowwdip} Density distribution in different slices
(color background) and 3D magnetic field lines (yellow lines) for
the case of an octupole plus weak dipole field with parameters
$\mu_1=0.2$, $\mu_3=0.3$, $\Theta_1=30^\circ$,
$\Theta_3=20^\circ$, $\phi_3=70^\circ$. Panel (a) shows an $xz$
slice at $t=0$; panels (b), (c), (d) show $xz$, $yz$ and $xy$
slices at $t=5$. The red lines show the Alfv\'{e}n surface, where
$\beta_1=1$. The thick cyan, white and orange lines represent the
rotation, dipole and octupole axes respectively.}
\end{figure}

\fig{flowwdip}(a) shows that initially, at $t=0$, the octupole
component dominates near the star, while the dipole dominates at
larger distances, such as the inner edge of the disc, which is
different from the pure octupole case. Panels (b)- (d) show that
the disc is stopped by the magnetosphere at $r_m\approx 0.9\approx
2.6 R_\star$. This radius is smaller than $r_{eq}\approx 1.5$, and
therefore the octupolar field dominates at the disc-magnetosphere
boundary. The accretion flow is similar to that in the pure
octupole case (see \fig{flowoct}), though here the misalignment
angle $\Theta=20^\circ$ is larger and matter flows more easily
above the equatorial set of loops. On the other hand, the octupole
is inclined more strongly (than in the pure octupole case), and
part of the funnel stream is located in the equatorial plane, due
to which the magnetospheric gap is not empty (panel d).
The red line in panel (d), $\beta_1=1$, shows that the
matter stress equals the magnetic stress at $r\approx 1$, where
the octupolar component dominates. At smaller accretion rates in the
disk, the line moves to larger distances, and at $r\gtrsim
1.5$, the dipole component dominates.

Comparison of panel (a) with the other panels shows the difference
between the initial potential field (panel a) and the
non-potential fields obtained in the simulations. The magnetic
field lines threading the disc and corona inflate, are wrapped
around the rotation axis and form some kind of a magnetic tower
(Romanova et al.\ 2011a). The field strongly differs from the
potential one at $r>r_m$, where the disc and corona matter
strongly disturb the magnetosphere. However, inside the
magnetospheric radius, $r<r_m$, the magnetic stress dominates, and
hence the magnetosphere is not disturbed by the plasma flow, and
the potential approximation is valid.

\begin{figure}
\centering
\includegraphics[scale=1.2]{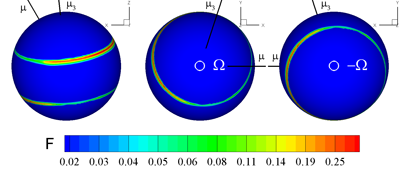}
\caption{\label{hswdip} The energy flux distribution on the
star with a strong octupole and weak dipole magnetic field,
$\mu_1=0.2$, $\mu_3=0.3$, $\Theta_1=30^\circ$,
$\Theta_3=20^\circ$, $\phi_3=70^\circ$, at $t=5$, as seen from the
equatorial plane (left panel), the north pole (middle panel), and
the south pole (right panel). }
\end{figure}

For the calculation of the light curves, we use the same approach
as Romanova et al.(2004). The total energy of the accreting matter
is assumed to be converted into isotropic blackbody radiation on
the surface of the star. The specific intensity of radiation from
a position $\bm{R}$ on the stellar surface into a solid angle
$\mathrm{d}\Omega$ in the direction $\bm{\hat{k}}$, is
$I(\bm{R,\hat{k}})=(1/\pi)F(\bm{R})$, where $F(\bm{R})$ is the
total energy flux of the inflowing matter,
$\theta=\arccos{({\bm{\hat R}}\cdot{\bm{\hat{k}}})}$. Therefore we
obtain the radiation energy received per unit time in the
direction $\bm{\hat{k}}$, $J=r^2F_{obs}=\int
I(\bm{R,\hat{k}})\cos\theta\mathrm{d}S$, where $r$ is the distance
between the star and the observer, $F_{obs}$ is the observed flux,
and $\mathrm{d}S$ is the surface area element. Simulations show
that, if the tilt of the dipole field is not very small
($\Theta\gtrsim 15^\circ$) and  the accretion rate does not vary
much with time, then the spots ``choose" their favorite positions
on the surface of the star, where they vary a small amount
(say, within $10^\circ$). \footnote{In the opposite case of a
smaller tilt and/or a highly variable accretion rate, a spot can
change its position significantly; it can even rotate with its own
angular velocity that is different from that of the star
(\citealt{roma04}, \citealt{bach10}).} This is why we fix the
spots at some moment of time and rotate the star to obtain the
light curves.


\fig{hswdip} shows the distribution of the energy flux,
$F$, on the surface of the star. The energy flux distribution is
usually very similar to that of the density distribution. One can
see that the spots in a purely octupolar case are similar to spots in
this case. However, now the rings have a higher inclination relative
to the equatorial plane due to the larger tilt of the octupolar
axis (compared with the purely octupolar case).

\begin{figure}
\centering
\includegraphics[scale=1.0]{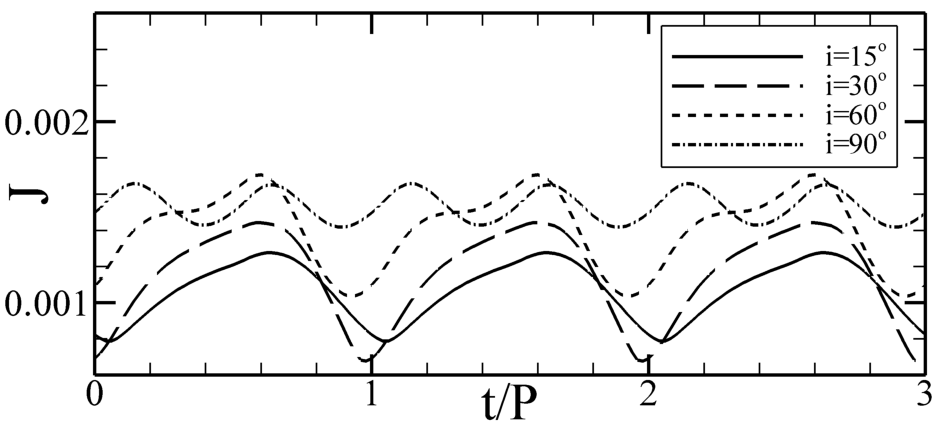}
\caption{\label{lcwdip} The light curves in the case of a strong
octupole and a weak dipole, $\mu_1=0.2$, $\mu_3=0.3$
and $\Theta_1=30^\circ$, $\Theta_3=20^\circ$, $\phi_3=70^\circ$,
for different inclination angles $i$. }
\end{figure}

\fig{lcwdip} shows the light curves associated with the rotation
of the star. The shapes of the light curves strongly depart from
sinusoids, in contrast with the pure octupolar case. However, the
cause is not the small dipole component, but instead the fact that
the misalignment angle ($\Theta_3=20^\circ$) is larger than that
in the pure octupolar case ($\Theta_3=10^\circ$), and a
part of the southern ring is seen by the observer even at small
inclination angles $i$. Each ring has a density
(brightness) enhancement connected with the tilt, and these bright
regions are in anti-phase (i.e., diametrically-opposite sides of the
star). Hence, at small $i$, the observer mainly sees one ring, while
only seeing part of the second ring, which gives one peak per period. At
large $i$, the observer sees both rings in similar proportion, but
the bright regions of rings alternate, and hence the light-curve
has two peaks per period. See the animations for spots viewed at
inclination angles $i=0^\circ$ and $i=90^\circ$ respectively.
\footnote{See Supplement: animations oct-dip-i0.avi and
oct-dip-i90.avi.}

The above results are shown for the accretion rate determined
in our code, in which the disk interacts with the octupolar component of
the field. If the accretion rate decreases, then the inner
disk will move to larger distances and the disk will start
interacting with the dipolar component. This will lead to a different
light-curve shape and to the phase shift in the
light-curve equal to the phase difference between the dipolar and
octupolar moments (see Sec. \ref{sec:phase} for details).

\subsection{Strong dipole and weak octupole}

We now consider the case where the dipole component ($\mu_1=2.0$) is much stronger than the octupole component
($\mu_3=0.2$). The misalignment angles are $\Theta_1=20^\circ$, $\Theta_3=10^\circ$, the phase angle is
$\phi_3=180^\circ$.

\begin{figure}
\centering
\includegraphics[scale=1.2]{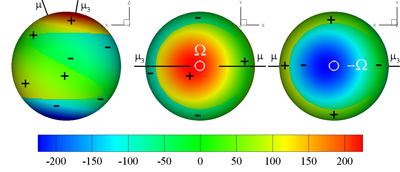}
\caption{\label{bsurfsdip} The surface magnetic field strength of
a star with a strong dipole and weak octupole magnetic field,
$\mu_1=2.0$, $\mu_3=0.2$, $\Theta_1=20^\circ$,
$\Theta_3=10^\circ$, $\phi_3=180^\circ$, as seen from the
equatorial plane (left panel), the north pole (middle panel) and
the south pole (right panel).}
\end{figure}

From \eqn{e-rcrit}, we find the critical radii where the dipole and
octupole components are equal: $r_{eq}=0.39$ and
$r_{pole}=0.45$. These radii are only slightly larger than the
radius of the star, $R_\star=0.35$, and hence the dipole
determines the dynamics of the flow in the entire vicinity of the star.

\fig{bsurfsdip} shows the magnetic field on the surface of the star.
The field distribution is complex and different
from a dipole or octupole field. There are two magnetic poles
with different polarities near the octupole axis, and two
belts of different polarities caused by the octupole component.
The field in the belts, however, is strongly distorted
by the dipole component.

\begin{figure}
\centering
\includegraphics[scale=1.1]{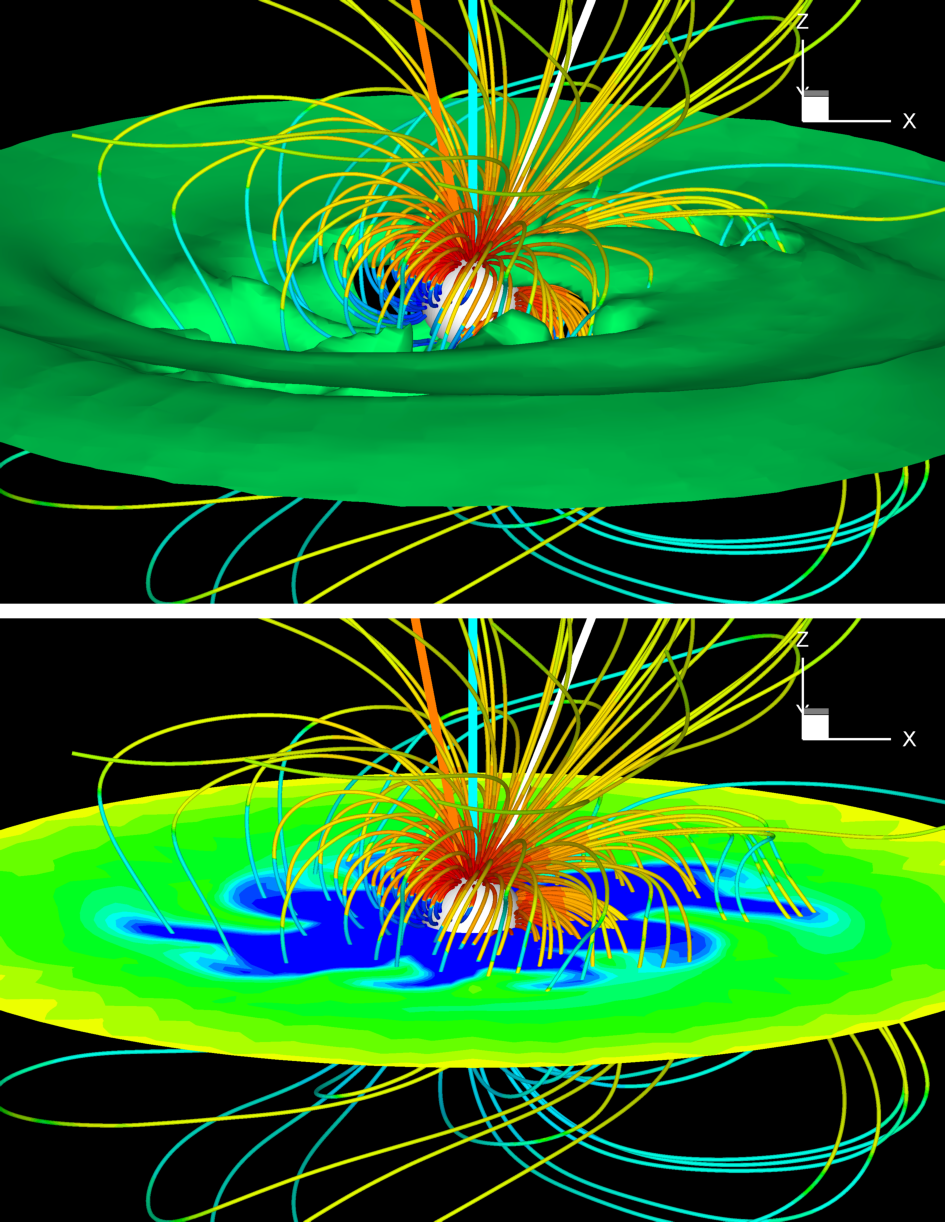}
\caption{\label{3dsdip} 3D views of matter flow to a star with a strong dipole and weak octupole magnetic field,
$\mu_1=2.0$, $\mu_3=0.2$, $\Theta_1=20^\circ$, $\Theta_3=10^\circ$, $\phi_3=180^\circ$, at $t=9$. The disc is shown by
a  constant density surface in green with $\rho=0.25$ in the top panel; different density levels in the disc plane are
shown in the bottom panel. The colors along the field lines represent different polarities and strengths of the field.
The thick cyan, white and orange lines represent the rotation, dipole and octupole axes respectively. }
\end{figure}

\fig{3dsdip} shows the 3D views of the accretion flow to the star
at $t=9$. The magnetic field looks like a dipole at all distances.
The matter is stopped by the dipole component of the magnetosphere
and is channelled to the polar regions in two funnel streams,
which is typical for accretion onto a star with a dipole field
(Romanova et al.\ 2003). In the disc plane (bottom panel), the
matter is stopped by the dipole-like field. One can see
that the matter in the disc penetrates through the external layers of
the magnetosphere due to the magnetic Rayleigh-Taylor instability
which we often observe in 3D simulations \citep{roma08,kulk08}.

\begin{figure}
\centering
\includegraphics[scale=1.0]{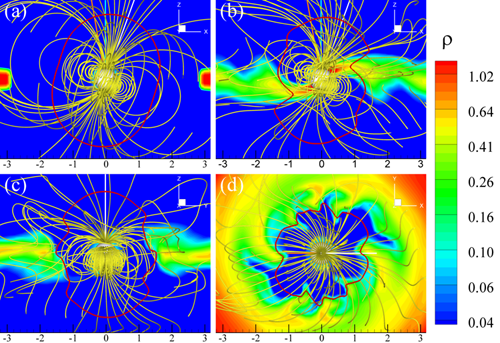}
\caption{\label{sdip} Density distribution in different slices
(color background) and 3D magnetic field lines (yellow lines) for
the case of a strong dipole and a very weak octupole field with
parameters $\mu_1=2.0$, $\mu_3=0.2$, $\Theta_1=20^\circ$,
$\Theta_3=10^\circ$, $\phi_3=180^\circ$. Panel (a) shows an $xz$
slice at $t=0$; panels (b), (c), (d) show $xz$, $yz$ and $xy$
slices at $t=9$. The red lines show the magnetospheric surface,
where $\beta_1=1$. The thick cyan, white and orange lines
represent the rotation, dipole and octupole axes respectively.}
\end{figure}

\fig{sdip} (panel a) shows that initially, at $t=0$, the magnetic
field has a dipole shape in the whole simulation region excluding
the close vicinity of the star. Panels (b) and (c) show that the
disc matter is stopped by the dipole-like magnetosphere and
accretes towards the poles in two funnel streams. Panel (d) shows
that a low-density magnetospheric gap forms around the star in the
disc plane. Comparison of panel (a) with panels (b)-(d) shows that
the potenti al approximation of the initially dipole magnetic
field shown in panel (a) stays valid at later times only inside
the magnetospheric surface, $r < r_m$, where $\beta_1<1$. At
larger distances, the field lines are dragged by the
disc and corona, and the field strongly departs from the initially potential dipole field. 

\begin{figure}
\centering
\includegraphics[scale=1.2]{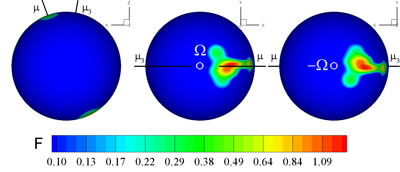}
\caption{\label{hssdip} The hot spots viewed from different angles for the case of $\mu_1=2.0$, $\mu_3=0.2$,
$\Theta_1=20^\circ$, $\Theta_3=10^\circ$, $\phi_3=180^\circ$: from the equatorial plane (left panel); from the north
pole (middle panel); and from the south pole (right panel). The color represents the density.}
\end{figure}

The hot spots are shown in \fig{hssdip}. As we expect, there are
only two polar spots, which is typical for accretion onto a star
with a dipole field. No ring-like spots are observed because the
dipole component dominates almost everywhere and matter is not
channeled by the octupole-like field. The hot spots are located
near the northern and southern magnetic poles. See our
supplemental animations for spots viewed at inclination angles
of $i=0^\circ$ and $i=90^\circ$, respectively. \footnote{See
Supplement: animations dip-oct-i0.avi and dip-oct-i90.avi.}

\begin{figure}
\centering
\includegraphics[scale=1.0]{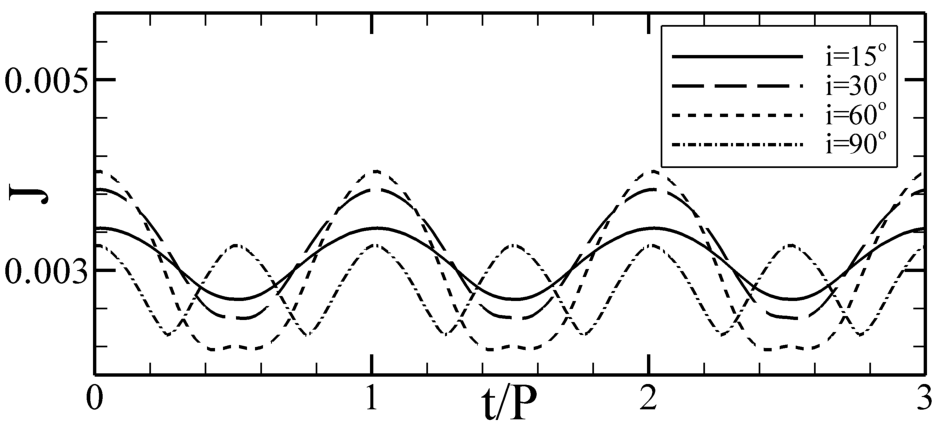}
\caption{\label{lcsdip} The light curves in the case of a strong dipole and weak octupole, $\mu_1=2.0$, $\mu_3=0.2$,
$\Theta_1=20^\circ$, $\Theta_3=10^\circ$, $\phi_3=180^\circ$, for different inclination angles $i$.}
\end{figure}

\fig{lcsdip} shows the light curves associated with the rotation
of the star at $t=9$. The shapes of the light curves are quite
sinusoidal for almost all inclination angles, which is also
typical for stars with slightly tilted dipole fields (Romanova et
al.\ 2004). At large $i$, two peaks per period are observed,
because the spot near the southern magnetic pole becomes visible
and contributes to the light curves.

\subsection{Dipole and octupole of comparable strength}

\begin{figure*}
\centering
\includegraphics[scale=1.3]{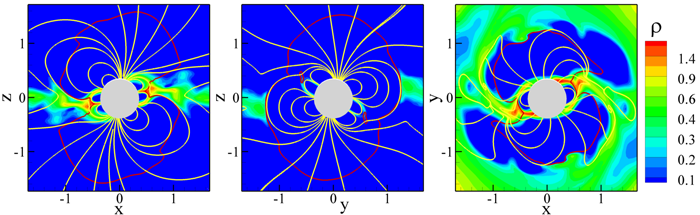}
\caption{\label{dip-oct} Close view of the density distribution in
different slices (color background) and magnetic field lines
(yellow lines) for the case of a strong dipole and octupole
magnetic field with parameters $\mu_1=1.0$, $\mu_3=0.3$,
$\Theta_1=30^\circ$, $\Theta_3=20^\circ$, $\phi_3=70^\circ$. From
left to right:  $xz$, $yz$ and $xy$ slices at $t=3.5$. The red
line shows the magnetospheric surface, where $\beta_1=1$.}
\end{figure*}

We now consider an interesting case where both dipole and octupole components disrupt the disc and channel matter. The
parameters are: $\mu_1=1$, $\mu_3=0.3$, $\Theta_1=30^\circ$, $\Theta_3=20^\circ$, $\phi_3=70^\circ$. For these
parameters we find from \eqn{e-rcrit}: $r_{eq}=0.67$, $r_{pole}=0.77$. However, the inclination of the dipole and
octupole axes could also influence the result. \fig{dip-oct} shows a close view of the matter flow. In the $xz$ plane,
the funnel streams are first channeled by the dipole-like field. When the matter flows close to the star, at about $2
R_\star$, the octupole determines the flow and converts each funnel stream into three small accretion streams between
the loops of field lines. In the $yz$ plane, the disc is stopped by the dipole component, and only a small amount of
matter flows along the dipole-like field lines and is then governed by the octupole. The $xy$ plane shows that matter
flow is strongly non-axisymmetric: in some directions the disc matter is stopped by the dipole, and in other directions matter flows much closer to the star and is stopped by the octupolar component.

\begin{figure}
\centering
\includegraphics[scale=1.2]{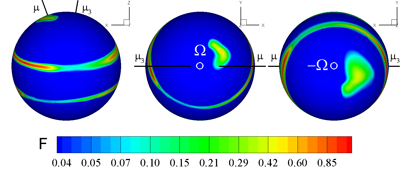}
\caption{\label{dip-oct-spot} The hot spots viewed from different angles for the case of $\mu_1=1.0$, $\mu_3=0.3$,
$\Theta_1=30^\circ$, $\Theta_3=20^\circ$, $\phi_3=70^\circ$: from the equatorial plane (left panel); from the north
pole (middle panel); and from the south pole (right panel). The color represents the density. }
\end{figure}

\fig{dip-oct-spot} shows the hot spots in this case. Most of the matter is governed by the octupolar field and forms
ring-like spots. Some matter flows into the polar regions located near the dipole magnetic poles.

\section{Phase shifts in light curves of stars with complex fields}
\label{sec:phase}

Here, we consider the possible mechanisms causing phase
shifts in the light-curves of stars with complex fields. We consider two main
mechanisms. First, if the magnetic field of the star reconstructs,
then this leads to a shift of the magnetic poles ; second, if the
accretion rate varies with time, then the disk interacts with
different multipoles, which have different phases. Below, we discuss
these two possibilities.


\subsection{Phase shift due to field reconstruction}
\label{sec:phase-reconstr}

 The magnetic field of a star may vary
with time due to internal dynamo processes. Such variations may
lead to a reconstruction of the field, and different multipoles
may dominate after the reconstruction. Or, the reconstructed field
may have the same multipoles, but with different orientations of
their magnetic axes. Here we investigate an example where a
predominantly octupolar field changes the orientation of its
magnetic axis. We choose an octupole with $\mu_3=0.3$, which in
state (a), has its magnetic axis at $\Theta_3=20^\circ$ and
$\phi_3=70^\circ$. Later, in state (b), it has new orientation
angles, $\Theta_3=60^\circ$ and $\phi_3=150^\circ$. In both cases
we have a small dipole component: $\mu_1=0.2$,
$\Theta_1=30^\circ$, which does not change.

\begin{figure}
\centering
\includegraphics[scale=1.0]{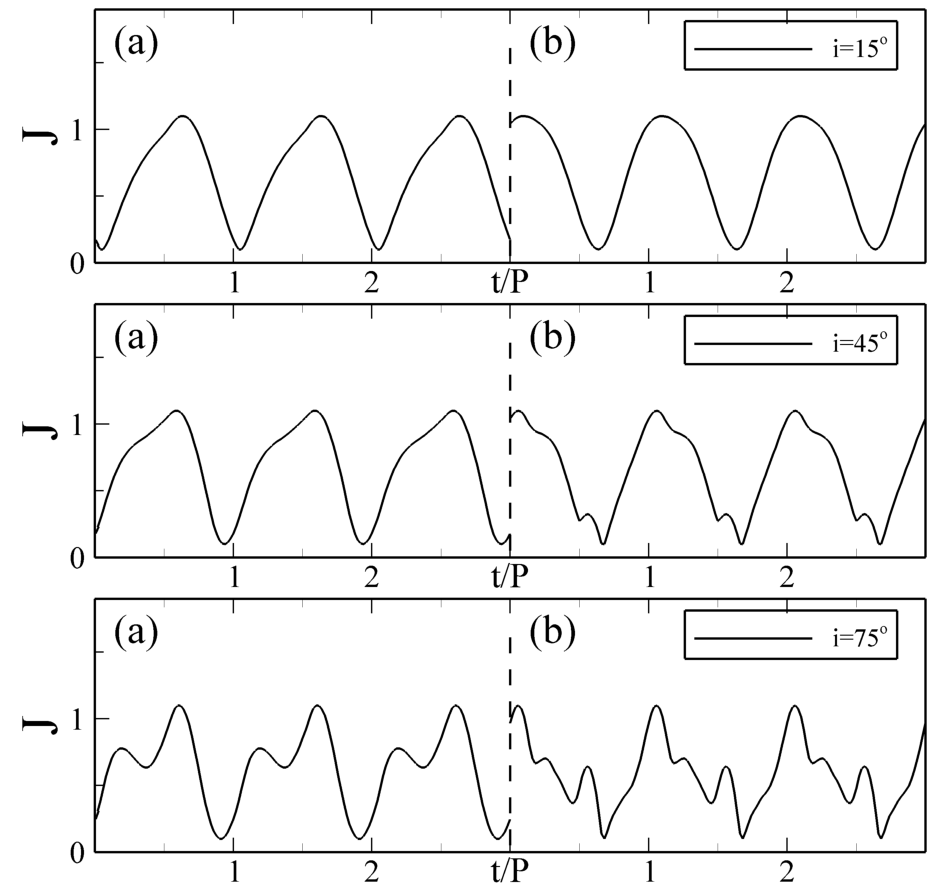}
\caption{\label{psreconstrucion} The phase shift due to the
magnetic field reconstruction. The field is predominantly the
octupolar field ($\mu_3=0.3$, $\mu_1=0.2$, $\Theta_1=30^\circ$),
and the octupolar axis changes its orientation. In the state (a)
$\Theta_3=20^\circ$ and $\phi_3=70^\circ$, in the state (b)
$\Theta_3=60^\circ$ and $\phi_3=150^\circ$. }
\end{figure}

We observe that in both cases the disc interacts with the
octupole-like field close to the star (like in \S 3.1) and the hot
spots represent two octupolar rings. However, the orientation of
the rings and their brightness distribution are different due to
the different axis directions. Most importantly, the octupole axes
have different directions relative to the meridional plane, which
leads to changes in hot spots and this, in turn, determines the
phase shift. \fig{psreconstrucion} shows that the light curves in
states (a) and (b) have different phases with a phase shift of
$\Delta\Phi=\Phi_b-\Phi_a=170^\circ$, where the amplitudes are
normalized to the same values. The shapes of the light curves are
different because the angle $\Theta_3$ is different in these
states, although this difference may not always lead to changes in
the pulse shapes.

In most cases it is not well known how fast the dynamo-driven
magnetic field reconstruction is. Observations of the CTTSs type
young stars show that possibly the field may vary on the scales of
months or even weeks (Smirnov et al.\ 2004; Donati et al.\ 2008).
What the time scales of magnetic field variations in neutron stars
and white dwarfs are is less clear.

We should mention that such phase shifts could occur during the
reconstruction of a field with any combination of multipoles,
including a purely dipole configuration.

\subsection{Phase shifts due to accretion rate variation}
\label{sec:phase-accr}

In another example we consider a star with a fixed complex field where the different multipoles are oriented at
different meridional angles, $\phi_n$. Then, at high accretion rates, the disc comes closer to the star and interacts
with the higher order multipoles, while at lower accretion rates, the truncation distance of the disc moves away from
the star and interacts with the lower order multipoles.

To study this case, we need a slight reinterpretation of our reference units defined in \S\ref{ref-units}. So far, we
have used the \textit{dimensionless} parameters $\mu_1$ and $\mu_3$ to control the magnetic moments of the star.
However, they can also be used to control the accretion rate instead, which is more convenient in this subsection; all
we need to do is recast the reference units into a different form. In order to do that, we note that the
\textit{dimensional} dipole moment of the star is $\mu_{1\star} = \mu_1\mu_{1,0}$. In the previous sections, we kept
the reference value $\mu_{1,0}$ fixed, which is why varying $\mu_1$ corresponded to varying the dipole moment. Now,
however, we keep $\mu_{1\star}$ fixed. It is then convenient to express $\mu_{1,0}$ in terms of $\mu_{1\star}$ as
$\mu_{1,0} = \mu_{1\star} / \mu_1$. Then, $\mu_{1,0} = B_0 R_0^3$ gives $B_0 = \mu_{1,0} / R_0^3 = \mu_{1\star} / \mu_1
R_0^3$, and $\mu_{3,0} = B_0 R_0^5$ becomes $\mu_{3,0} = \mu_{1\star} R_0^2/ \mu_1$. The dimensional octupole moment
then is $\mu_{3\star} = \mu_3\mu_{3,0} = (\mu_3/\mu_1) \mu_{1\star} R_0^2$. The reference accretion rate is $\dot M_0 =
\rho_0 v_0 R_0^2 = B_0^2 R_0^{5/2}/\sqrt{GM} = \mu_{1\star}^2 / (\mu_1^2 R_0^{7/2}\sqrt{GM})$. The dimensional
accretion rate $\dot M_{dim}$ is then given by
\begin{equation}
\label{mdot} \dot M_{dim} \approx \left(\frac{\dot M}{\mu_1^2}\right)\frac{\mu_{1\star}^2}{(GM_\star)^{1/2} R_0^{7/2}}.
\end{equation}
It is now clear that changing the \textit{dimensionless} parameter $\mu_1$ has the effect of changing the accretion
rate $\dot M_{dim}$, while keeping $\mu_{1\star}$ fixed as mentioned above. To keep $\mu_{3\star}$ fixed as well, we
simply have to change $\mu_3$ such that the ratio $\mu_{3\star} / \mu_{1\star}$ is fixed. This allows us to change the
accretion rate while keeping the stellar magnetic field fixed.

The physical meaning of this recasting is the following. The most natural interpretation of changing $\mu_1$ and
$\mu_3$ is changing the stellar magnetic field, the result of which is a decrease in the magnetospheric radius.
However, we can also decrease the magnetospheric radius by fixing the stellar magnetic field and increasing the
accretion rate instead, which is exactly what \eqn{mdot} shows. The dimensionless parameters $\mu_1$ and $\mu_3$ are
therefore best thought of as controlling the {\it magnetospheric size}.

This yields us the convenient possibility of using $\mu_1$ to
represent two states of a star with different accretion rates to
investigate the phase shifts between them. As an example, we take
a star with a superposition of dipole and octupole fields, where
they have parameters, $\Theta_1=30^\circ$, $\Theta_3=20^\circ$,
$\phi_3=70^\circ$ and fixed ratio of $\mu_3/\mu_1$. It is
important that the dipole and octupole have different phase angles
$\phi_n$.

\begin{figure}
\centering
\includegraphics{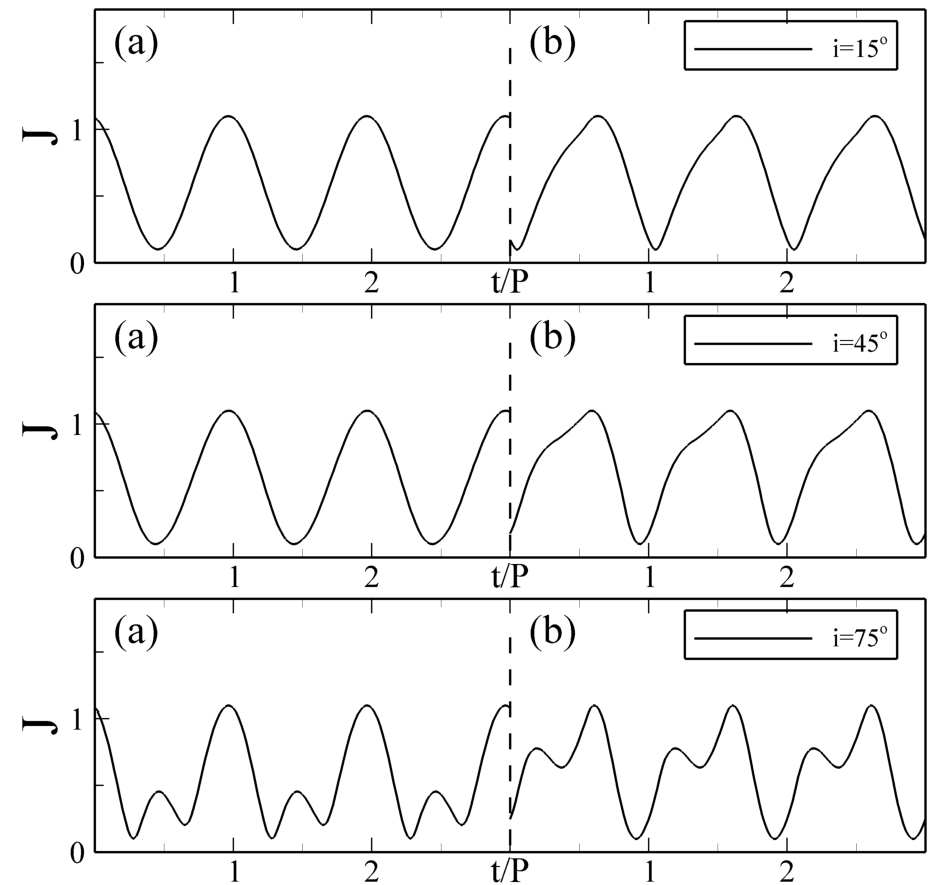}
\caption{\label{psmdot} The phase shift due to the variation of
the accretion rate. (a) For low accretion rate ($\mu_1=2$), the
dipole component stops the disc. (b) For high accretion rate
($\mu_1=0.2$), the octupole component stops the disc. The phase
shift occurs at the transition between these two states.}
\end{figure}

We choose two states with the above parameters for the dipole and
octupole, but with different $\mu_1$: state (a) with $\mu_1=2$;
and state (b) with $\mu_1=0.2$. We observe from simulations that
in state (a) the disc  interacts mainly with the dipole component,
while in state (b) it  interacts mainly with the octupole
component. \fig{psmdot} shows a phase shift of peaks
$\Delta\Phi=\Phi_b-\Phi_a=120^\circ$ between the states (a) and
(b), where the amplitudes are normalized to the same values. The
phase shift occurs because the octupole and dipole axes are
located in different meridional planes and have different angles
$\phi$, due to which the matter flows to different places of the
star at different accretion rates.

This type of phase shift is expected during periods of accretion
rate variation. What is the accretion rate ``jump" between states
(a) and (b)? We can give a simple estimate, $(\dot M)_b/(\dot
M)_a=((\mu_1)_a/(\mu_1)_b)^2=(2/0.2)^2=100$. However, we should
notice that in state (b) the octupole strongly dominates and
\eqn{mdot} may not be used directly. In addition, the transition
may occur at smaller values of $\mu_1$ in state (a). So it is
clear that this type of phase shift may be produced by accretion
rate (luminosity) variation.

Another point that should be noted is that the phase shifts would
be expected to be accompanied by changes in the light curve pulse
profile, because the shape and position of the hot spots are very
different at different field configurations.

\section{Discussion of Phase Shifts in Accreting Millisecond Pulsars (AMPs)}
\label{sec:timing-noise}

Different types of phase shifts are observed in accreting
millisecond pulsars,  in which a weak neutron star magnetic field
($B\sim 10^8-10^9$ G) truncates the accretion disk at a distance
of a few stellar radii, and in which the disk-magnetosphere
interaction determines many of observational properties of the
neutron star (e.g., Wijnands \& van der Klis 1998; Chakrabartiy \&
Morgan 1998). The main type of phase shift is connected with
so-called \textit{timing noise} which appears as irregular
pulse-phase jumps on timescales from hours to months (e.g.,
Burderi et al. 2006; Hartman et al. 2008). The amplitude of
pulse-phase offsets usually correlated with X-ray flux (e.g.,
Papitto et al. 2007; Reggio et al. 2008; Patruno et al. 2009a,b)
though sudden strong jumps in phase are also observed (e.g.,
Burderi et al. 2006; Hartman et al. 2008, 2009). This phenomenon
is not completely understood yet. The correlation of the
pulse-phase offsets with X-ray flux may be an argument in favor of
a ``moving spots'' model, in which the position of hot spots on
the stellar surface depends on the accretion rate (Patruno et al.
2009a,b). Spot motions have been observed in 3D MHD simulations
(Romanova et al. 2003, 2004) and have been proposed as an
explanation of timing noise by Lamb et al. (2009a,b). This
explanation is a possibility. Alternatively, we suggest that the
phase shifts could be connected with the variation of the
accretion rate for a star with a complex field, as discussed in
Sec. \ref{sec:phase-accr}. In this model, the phase shift is also
proportional to the accretion rate, as in the moving spots model.
More detailed analysis is required to decide between these two
models.

The pulse shape usually varies during an outburst, and
hence the contribution of different harmonics varies in time. This
also leads to the timing noise (e.g., Poutanen et al. 2009;
Gierlinsky \& Poutanen 2005; Poutanen \& Gierlinsky 2003).
Moreover, there is a phase shift between the soft and hard X-ray
bands, in which the pulse arrival in the soft X-ray band is
lagging behind that in the hard X-ray band (Cui et al. 1998). This
phenomenon can be connected with the details of radiation in the
shock near the surface of the neutron star \citep{pout03}, and it
adds complexity to the problem of timing noise \citep{gier05}. In
our model of ``complex fields + varying accretion rate'', the
pulse shape also varies in time, which also might contribute to
the timing noise. However, in our model, the phase jumps are
usually not at the origin of pulse shape changes. This is
different from the model proposed by, e.g., Poutanen et al.
(2009), which is focused on the pulse profile changes to
investigate the origin of timing noise in accreting pulsars.

In another phase-shift model proposed in this paper (see
Sec. \ref{sec:phase-reconstr}), the phase shifts are connected
with the magnetic field reconstruction due to the internal
processes in the star. In this case, the phase shifts (or jumps)
are expected to be random and occur with no correlation with the
X-ray flux.  This mechanism is definitely expected in some
accreting young solar-type stars, in which the magnetic field
varies rapidly due to convection processes inside the star (e.g.,
Donati et al. 2010). It is less clear whether this mechanism
operates in accreting neutron stars. The magnetic field of
\textit{isolated} millisecond pulsars varies only slowly in time
due to slow processes in the liquid core of the neutron star
(e.g., Ruderman 1991) or due to slow diffusion of the earlier
buried magnetic field through the crust of the neutron star (e.g.,
Cumming, Zweibel \& Bildsten 2001). In both cases, it is expected
that the global field of millisecond pulsars varies on time-scales
that are much longer than the observed bursts. However, the
situation can be different in cases of presently-accreting
millisecond pulsars, in which temporarily enhanced accretion
events can lead to temporary burial of parts of the stellar
magnetic flux, which can diffuse back much faster. Accretion
through the funnel streams leads to enhanced accretion towards the
magnetic poles (Lovelace, Romanova \& Bisnovatyi-Kogan 2005) and
to formation of temporary polar `mountains' at the poles (Payne \&
Melatos 2004, 2006), while the magnetic flux is pushed towards the
equator. This field will be redistributed at later times, after a
thermonuclear burst. Another possibility for the temporary
redistribution of the field has been observed in global
simulations of disk-magnetosphere interactions, in which the
accretion disk reaches the surface of the star during periods of
high accretion rate. Simulations show that, in cases of very high
accretion rates, the disk matter accretes through the equatorial
boundary layer, and it pushes parts of the field towards the
stellar surface \footnote{Global simulations show that the
magnetic field is strongly modified during equatorial accretion;
it looks like a a quadrupolar or more complex field (see, e.g.,
fig. 15 of \citealt{roma11b})}. We suggest that this field can be
temporarily buried by the accreting matter, and then it will
diffuse out relatively fast. Such a situation can be realized
during the maximum point of the outburst, which is when the disk
can reach the surface of the star. Subsequently, the field will
reconstruct relatively fast due to diffusion through a thin layer
of accreted matter. Reconstruction of this field will lead to the
phase jumps which would be observed on the background of the
gradually declining X-ray flux. Of course, this possibility should
be investigated further.

We should also note that in the model where the phase
varies due to variation of the accretion rate, the phase is
expected to return back to its original value, as soon as the
accretion rate returns back. However, in the model where the field
reconstructs in time, the phase will not return back.

\section{Conclusions and discussion}

We performed global 3D MHD simulations of accretion onto stars
with predominantly octupolar magnetic fields, and for a
combination of dipole and octupole fields. The calculation became
possible due to the enhancement of the grid resolution near the
star. Simulations have shown that:

\begin{enumerate}

\item

If the disc interacts predominantly with the octupolar field, then
matter flows into two octupolar rings on the surface of the star.
In the case of an inclined octupole field, the hot spots are
inclined and have a brightness amplification on one side of the
ring.

At small tilt angles of the octupole, $\Theta=10^\circ$, the light
curves are quite sinusoidal and ``mimic'' the shape of the light
curves of the pure dipole field, in particular at small
inclination angles $i$. However, at larger  $\Theta$ (e.g.,
$\Theta=20^\circ$), the light curves strongly depart from
sinusoidal, even for small $i$.

\item

If the dipole component strongly dominates, then matter flows in two funnel streams governed by the dipole field, and
two round spots form in the vicinity of the dipole magnetic poles. If the dipole and octupole have comparable strengths
at the truncation radius, then both components channel matter to the star. Usually both octupolar rings and polar spots
form on the surface of the star.

\item

The potential field approximation is valid only in the region
around the star where the magnetic stresses are higher than the
matter stresses. At larger distances, the field is dragged by the
disc and corona and strongly departs from being potential. The
external magnetic field usually acquires a significant azimuthal
component and inflates. A number of field lines wrap around the
rotation axis, forming a magnetic tower which may propagate to
larger distances due to magnetic force (e.g., Romanova et al.\
2011a).

\item

Accretion onto stars with multipolar fields may lead to phase
shifts in the light curves. This may occur in the
following situations:~
 ~ (a) If the complex magnetic field is fixed but the accretion rate
 varies, then the disk interacts with multipoles of
 different orders that may have different tilts, $\theta_n$, and phase angles,
 $\phi_n$. These phase-shifts will correlate with the variation of the accretion rate.~
(b) If the magnetic field of the star varies in time due
to, e.g., dynamo process (in young stars) or due to some other
processes (e.g., temporary burial of the field in compact stars), then
the phase-shifts will be connected with
reconstruction of the field and motion of magnetic poles. It is
expected that the pulse profiles will vary during these phase-shifts,
because the shapes and positions of the hot spots are
different in different field configurations.

\end{enumerate}

These new challenging 3D simulations helped us to understand the
nature of accretion onto stars with octupolar fields. Recent
measurements by Donati et al.\ (2007, 2008) of the two CTTSs V2129
Oph and BP Tau have shown that their fields have a significant
octupolar component. In our next papers (Romanova et al. 2011a;
Long et al. 2011) we compare our numerical model with observations
of these stars.

Magnetic fields in young, T Tauri-type stars, may have a complex
structure and may vary with time due to internal dynamo processes.
The recently measured magnetic fields of CTTS stars CV Cha and CR
Cha (Hussain et al.\ 2009) show a complex structure consisting of
a number of multipoles. In another CTTS, V2247 Oph, the magnetic
field is complex and varies on the time-scale of weeks (Donati et
al. 2010; see also Smirnov et al. 2005; Donati et al. 2011).
Frequent phase shifts are expected in this star due to internal
field reconstruction, as discussed in this paper. In all these
stars the phase shifts may also occur due to accretion rate
variations and interactions with different multipoles.

Photometric variability of CTTSs show complex patterns and
are still puzzling  (e.g., Herbst et al. 1994; Percy, Gryc \& Wong
2006; Grankin et al. 2007). It is often the case that, instead of
a definite period, a quasi-period is observed, and its frequency
varies with time (e.g., Alencar \& Batalha 2002; Rucinski et al.
2008). This may be connected with the phase-shift of the main
period, in which the period varies on a time-scale of the rotation
of the star. This phenomenon of phase-shifts should be further
investigated.b

Phase-shifts are observed in millisecond pulsars, and
we discuss possible application of our models to these objects in
Sec. \ref{sec:timing-noise}. Similar phase-shifts are expected in
accreting, weakly-magnetized white dwarfs(e.g., Dwarf Novae or
in Intermediate Polars).


\section*{Acknowledgments}

This research was conducted using the NASA High End Computing
Program computing systems, specifically the Columbia, Discover and
Pleiades superclusters. The authors thank A.V. Koldoba and G.V.
Ustyugova for the earlier development of the codes, and A.A.
Blinova, A.K. Kulkarni and R.V.E. Lovelace for helpful
discussions. The research is supported by NSF grant AST0709015,
and funds of the Fortner Endowed Chair at Illinois. Research of
MMR is supported by NASA grant NNX08AH25G and NSF grant
AST-0807129.


\appendix

\section{The magnetospheric radius}

The magnetospheric radius $r_m$ is the radius where the disc is truncated by the magnetosphere.
It is also used to depict the characteristic size of the magnetosphere. Here we estimate
the magnetospheric radii for various multipoles. For simplicity we assume that they
are aligned with the rotation axis.

\subsection{The Magnetospheric Radius for A Dipole Field}\label{rm-dip}

Here we derive the general formula for the truncation (magnetospheric) radius $r_m$ in the cases of the dipole fields.
The disc is truncated where
the magnetic stresses equal the matter stresses,
$B^2/8\pi=p+\rho v_\phi^2$. We
assume that the disc is cold,  $p<<\rho v_\phi^2$,
hence $B^2/8\pi=\rho v_\phi^2$.

We also assume that the system is in a stationary state and matter is supplied from the viscous disc with
a constant accretion rate $\dot M =4\pi
rhv_r\rho$, where $r$ is the distance to the star, $h=(c_s/v_K)r$ is the disc scale height, $v_r=\alpha c_s^2/v_K$ is
the radial flow velocity, $c_s$
is the sound speed, $v_K=(GM/r)^{1/2}$ is the Keplerian velocity, $\alpha$ is the Shakura-Sunyaev
viscosity parameter. We obtain,
\begin{equation}\label{A1}
\rho v_\phi^2\approx\rho v_K^2=(4\pi\alpha)^{-1}(v_k/c_s)^3\dot{M}(GM)^{1/2}r^{-5/2}.
\end{equation}

For the equatorial component of the {\it dipole field} the magnetic field is $B=\mu_1/r^3$.
Substituting into the equation $B^2/8\pi=\rho v_\phi^2$
 and using \eqn{A1}, we obtain,
\[r_m = k_1 r_{m}^{(0)},~~ k_1=(\alpha^2/2)^{1/7}(c_s/v_K)^{6/7},\]
\begin{equation}\label{A2}
r_{m}^{(0)} = \mu_1^{4/7}\dot{M}^{-2/7}(GM)^{-1/7}.
\end{equation}
The main term, $r_m^{(0)}= \mu_1^{4/7}\dot{M}^{-2/7}(GM)^{-1/7}$, exactly coincides with the main term for spherical
accretion (see also Bessolaz et al.\ 2008). However, here the coefficient $k_1$ depends on $\alpha$ and the ratio
$c_s/v_K$. We should note that the formula for $\dot M$ used in this derivation is applicable only for thin discs, and
is only approximately valid near the magnetosphere, where the disc matter is accumulated, the disc becomes thicker and
has an excess of matter pressure. That is why the coefficient $k_1$ can be considered to be approximate and can be
substituted with some number. Comparison of numerical results for $r_m$ with  \eqn{A2} made in Long et al.\ (2005)
shows that for their set of simulations, $k_1\approx 0.5$ (see more comparisons in Bessolaz et al.\ 2008).

\subsection{The Magnetospheric Radius for A Multipolar Field} \label{rm-mult}

Next, we  derive the magnetospheric radius for multipolar field. For simplicity, we assume that the main axis of the
multipolar field is {\it aligned} with the rotation axis, and therefore the $n-$th multipolar component of the magnetic
field can be presented in the form $B_n\sim\mu_n/r^{n+2}$. Similarly, we obtain,
\[r_{m,n}=k_n r_{m,n}^{(0)}, ~~ k_n=(\alpha^2/2)^{\frac{1}{4n+3}}(c_s/v_K)^{\frac{6}{4n+3}},\]
\begin{equation}\label{A3}
r_{m,n}^{(0)}=\mu_n^{\frac{4}{4n+3}}\dot{M}^{-\frac{2}{4n+3}}(GM)^{-\frac{1}{4n+3}}.
\end{equation}
This formula can be applied in the cases of aligned multipoles, or can be used for estimates of the magnetospheric
radius in general, misaligned cases.

We should note that the $B_z$ equals 0 in the disc plane for aligned fields of $2n$-th order multipoles, such as
quadrupole (ref. Long et al. 2007), and matter could flow directly to the star in the disc plane. So $r_{m,n}$ does not
reflect where the inflowing matter stops, but only the size of magnetosphere.


\bibliographystyle{elsarticle-harv}
\bibliography{<your-bib-database>}



\end{document}